\documentclass[aps,prl,twocolumn,superscriptaddress,showpacs]{revtex4}
\usepackage{amsfonts,amscd,amsmath, amssymb,graphicx,color}
\def\ep{\text{e}}

\def\g{\mathfrak{g}}
\def\s{\mathfrak{s}}
\def\c{\mathfrak{c}}
\def\4{\tfrac{1}{4}}

\def\zt{z_{\text{\tiny T}}}

\def\T{T_c}
\begin{document}
\preprint{SPAG-A1/09}
\title{Renormalized Polyakov Loop in the Deconfined Phase of SU(N) Gauge Theory and Gauge/String Duality}
\author{Oleg Andreev}
%\email[]{}
%\homepage[]{Your web page}
%\thanks{}
%\altaffiliation{}
\affiliation{Technische Universit\" at M\" unchen, Excellence Cluster, Boltzmannstrasse 2, 85748 Garching, Germany}
\affiliation{L.D. Landau Institute for Theoretical Physics, Kosygina 2, 119334 Moscow, Russia}
%\date{\today}
\begin{abstract}
We use gauge/string duality to {\it analytically} evaluate the renormalized Polyakov loop in pure Yang-Mills theories. For $SU(3)$, the 
result is in a quite good agreement with lattice simulations for a broad temperature range. 
\end{abstract}
\maketitle
%__________________________  I N T R O ______________________________

\section{Introduction}

It is well known that a pure $SU(N)$ gauge theory at high temperature undergoes a phase transition. This phase transition is of special interest
because of many its aspects can be characterized precisely \cite{pol}. In particular, the order parameter is given by the Polyakov loop

\begin{equation}\label{P1}
L(T)=\frac{1}{N}\text{tr Pexp}\Bigl[ig\int_0^{1/T}dt\, A_0 \Bigr]
\,,
\end{equation}
where the trace is over the fundamental representation, $t$ is a periodic variable of period $1/T$, with $T$ the temperature, $g$ is a 
gauge coupling constant, and $A_0$ is a vector potential in the time direction. The usual interpretation of \eqref{P1} is as a phase factor
associated to the propagation of an infinitely heavy test quark in the fundamental representation of the gauge group.

Until recently, the lattice formulation, still struggling with limitations and system errors, and effective field theories were the main 
computational tools to deal with non-weakly coupled gauge theories. The Polyakov loop was also intensively studied (see, for example, 
\cite{pis-rev} and references therein). The situation changed drastically with the invention of the AdS/CFT correspondence \cite{malda1} 
that resumed interest in another tool, string theory.

In this note we continue a series of recent studies \cite{az1,az2,a-pis} devoted to a search for an effective string description of pure gauge 
theories. In \cite{az1}, the model was presented for computing the heavy quark and multi-quark potentials at zero temperature. 
Subsequent comparison \cite{white} with the available lattice data has made it clear that the model should be taken seriously. Later, 
this model was extended to finite temperature. The results obtained for the spatial string tension \cite{az2} and the thermodynamics \cite{a-pis} 
are remarkably consistent with the lattice, too. As is known, QCD is a very rich theory supposed to describe the whole spectrum of strong 
interaction phenomena. The question naturally arises: How well does the model describe other aspects of quenched QCD? Here, we attempt to 
{\it analytically} evaluate the Polyakov loop as an important step toward answering this question \cite{az3}. In addition, a good motivation for this 
test is lattice data revealed recently by \cite{gupta}.

Before proceeding to the detailed analysis, let us set the basic framework. As in \cite{az1,az2,a-pis}, we take the following ansatz
for the five-dimensional background geometry

\begin{multline}\label{metric}
ds^2=G_{nm}dX^ndX^m=
R^2 w
\left(f dt^2+d\vec{x}^2+\frac{1}{f}dz^2\right)
\,,\\
w(z)=\frac{\ep^{\s z^2}}{z^2}
\,,\quad
f(z)=1-\bigl(\tfrac{z}{\zt}\bigr)^4
\,,\phantom{=\frac{\ep^{\s z^2}}{z^2}}
\end{multline}
where $\zt =1/\pi T$. $\s$ is a deformation parameter whose value can be fixed from the critical temperature \cite{s}. We take a constant 
dilaton and discard other background fields.

In discussing the Wilson and Polyakov loops within the gauge/string duality \cite{lit}, one first chooses a contour ${\cal C}$ on a four-manifold 
which is the boundary of a five-dimensional manifold. Next, one has to study fundamental strings on this manifold such that the string 
world-sheet has ${\cal C}$ as its boundary. In the case of interest, ${\cal C}$ is an interval between $0$ and $1/T$ on the $t$-axis. The 
expectation value of the Polyakov loop is schematically given by the world-sheet path integral

\begin{equation}\label{pol}
\langle\,L(T)\,\rangle=\int DX\,\ep^{-S_w}
\,,
\end{equation}
where $X$ denotes a set of world-sheet fields. $S_w$ is a world-sheet action. In principle, the integral \eqref{pol} can be evaluated approximately in
terms of minimal surfaces that obey the boundary conditions. The result is written as $\langle\,L(T)\,\rangle=\sum_n w_n\exp[-S_n]$, where 
$S_n$ means a renormalized minimal area whose weight is $w_n$.

%_______________________________________________________________________________________________

\section{Calculating the Polyakov Loop}

Given the background metric, we can attempt to calculate the expectation value of the Polyakov loop by 
using the Nambu-Goto action for $S_w$ in \eqref{pol}

\begin{equation}\label{ng0}
S=\frac{1}{2\pi\alpha'}\int d^2\xi\,\sqrt{\det \, G_{nm}^{}\partial_\alpha X^n\partial_\beta X^m\vphantom{\bigl(\bigr)}}
\,.
\end{equation}
Here $G_{nm}$ is the background metric \eqref{metric}. In the case of interest, this action describes a fundamental string stretched between 
the test quark on ${\cal C}$ (at $z=0$) and the horizon at $z=\zt$. Since we are interested in static configurations, we 
choose $\xi_1=t$, $\xi_2=z$. This yields

\begin{equation}\label{2ng}
S=\frac{\g }{\pi T}\int^{\zt}_0
dz\,w
\sqrt{1+f (\vec{x}\,')^2}
\,,
\end{equation}
where $\g=\frac{R^2}{2\alpha'}$. A prime stands for a derivative with respect to $z$.

Now it is easy to find the equation of motion for $\vec{x}$

\begin{equation}\label{eq-x}
\biggl[w f\vec{x}\,'/\sqrt{1+f(\vec{x}\,')^2}\biggr]'=0
\,.
\end{equation}
It is obvious that Eq.\eqref{eq-x} has a special solution $\vec{x}=const$ that represents a straight string stretched between 
the boundary and the horizon. Since this solution makes the dominant contribution, as seen from the integrand in \eqref{2ng}, we 
won't dwell on other solutions here.

Having found the solution, we can now compute the corresponding minimal area. Since the integral \eqref{2ng} is divergent at
$z=0$ due to the factor $z^{-2}$ in the metric, we regularize it by imposing a cutoff $\epsilon$

\begin{equation}\label{S}
S_{\text{\tiny R}}=\frac{\g}{\pi T}\int^{\zt}_ \epsilon dz\,w
\,.
\end{equation}
Subtracting the $\frac{1}{\epsilon}$ term (quark mass) and letting $\epsilon=0$, we get a renormalized area
\begin{equation}\label{S3}
S_0=\frac{\g}{\pi T}\int^{\zt}_ 0 dz\,\Bigl( w-\frac{1}{z^2}\Bigr)+c
\,,
\end{equation}
where $c$ is a normalization constant which is scheme-dependent.

Next, we can perform the integral over $z$. The result is 

\begin{equation}\label{S4}
S_0=\g\biggl(\sqrt{\pi}\frac{\T}{T}\text{Erfi}\Bigl(\frac{\T}{T}\Bigr)+1-\ep^{(\T/T)^2}\biggr)+c
\,.
\end{equation}
In this formula $\T$ is given by $\T=\sqrt{\s}/\pi$ \cite{az2}.

Combining the weight factor with the normalization constant as $\c=\ln w_0-c$, we find

\begin{equation}\label{P2}
L(T)=\exp\biggl[\c
-\g\biggl(\sqrt{\pi}\frac{\T}{T}\text{Erfi}\Bigl(\frac{\T}{T}\Bigr)+1-\ep^{(\T/T)^2}\biggr)
\biggr]
\,,
\end{equation}
with $\text{Erfi}(z)$ the imaginary error function. This is our main result. 

%_________________________________________________________________________________________
\section{Numerical Results and Phenomenological Prospects}

It is of great interest to compare the temperature dependence of \eqref{P2} with other results for the high temperature phase of $SU(N)$ 
gauge theory. In doing so, we start with lattice QCD. Clearly, $N=3$ is of primary importance. In Fig.1 a comparison is shown with the 
recent data of \cite{gupta}. We see that our model is in a quite good agreement with the lattice for a broad temperature 
range  $1.05\,\T \lesssim  T\lesssim 20\,\T$. The maximum discrepancy occurred at $T=1.05\,\T$ is of order 15\%. It rapidly decreases 
with temperature reaching 2\% at $T=2.2\,\T$ and  becoming almost negligible up to $20\,\T$. Then, it starts to grow back again. 

For completeness, we can fit the value of $\g$ to be $0.72$ that significantly improves  
%____________________________________  f -1 ______________________________________________
\begin{figure}[ht]
\centering
\includegraphics[width=6.3cm]{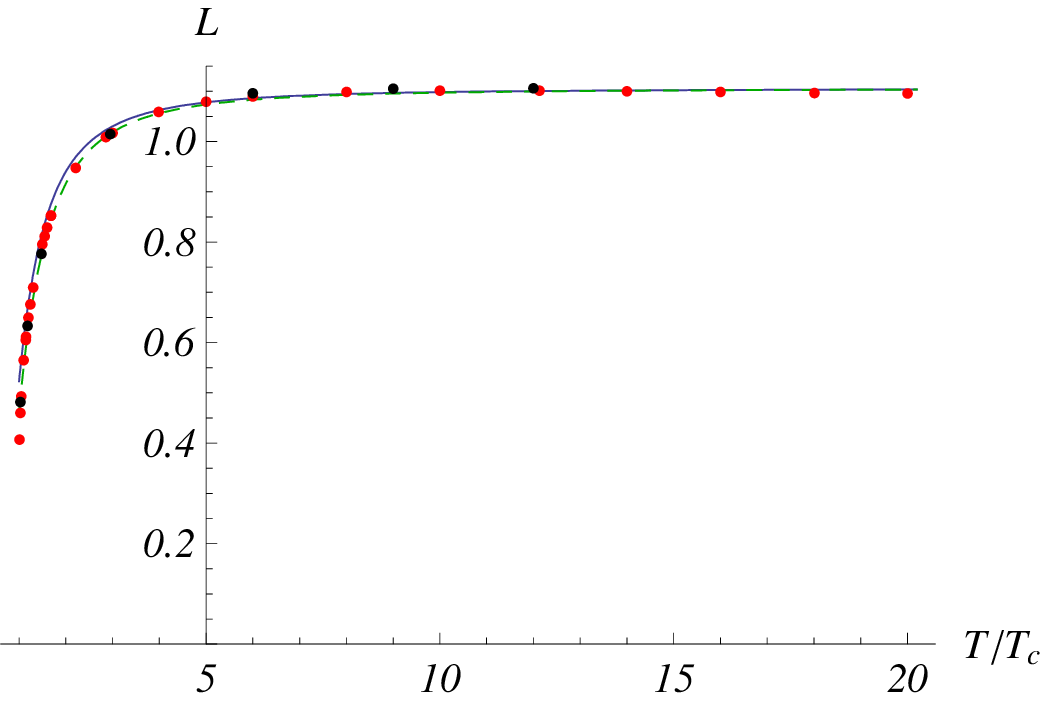}
\hfill
\includegraphics[width=6.3cm]{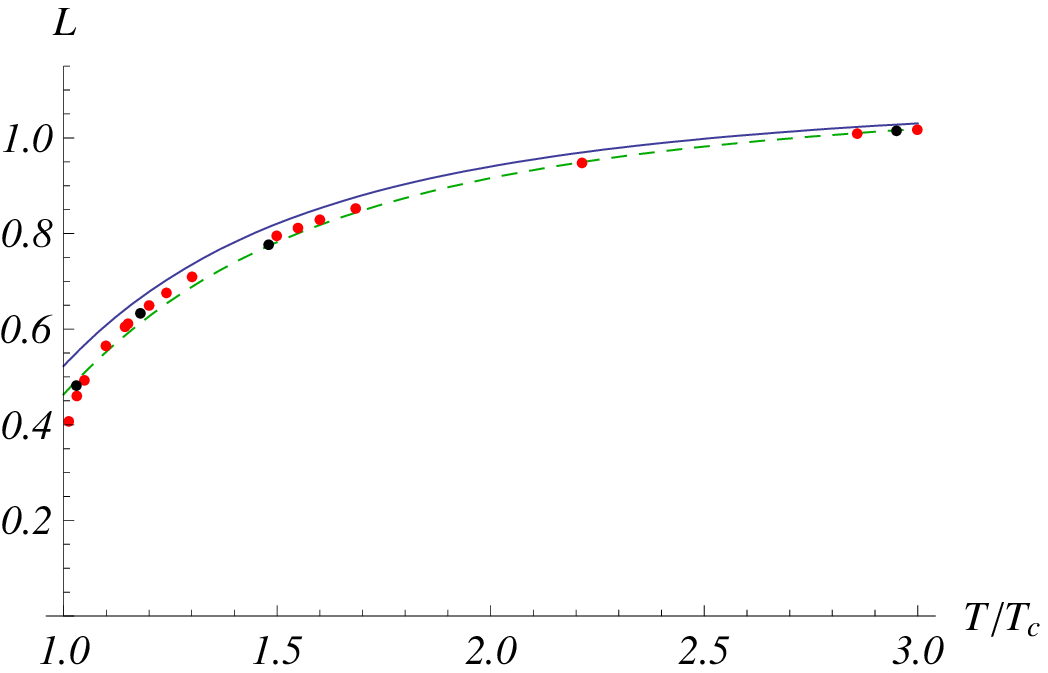}
\caption{\small{ The renormalized Polyakov loop in $SU(3)$ gauge theory. The solid blue curve 
corresponds to \eqref{P2} with $\g=0.62$ as fixed from the heavy quark potential at zero T in \cite{white}. The dashed green curve represents 
the "best fit'' with $\g=0.72$. In both cases, the value of $\c$ is set to $0.10$. The dots are from lattice simulations of \cite{gupta}. The red dots 
are for $N_\tau=4$, while the black dots are for $N_\tau=8$. We do not display any error bars because they are quite small, comparable to 
the size of the symbols. }}
\end{figure}
%________________________________________________________________________________________
accuracy. For example,  at $T=1.05\,\T$ it becomes of order 6\%. One possible explanation for the better fit is that we have 
evaluated \eqref{pol} classically (in terms of strings). If we take into account semi-classical corrections, then the value of $\g$ gets 
renormalized.

For practical purposes, the expression \eqref{P2} looks somewhat awkward. Following \cite{a-pis}, we expand $S_0$ and $L$ in powers of 
$(\T/T)^2$. If we ignore all higher terms, then a final result can be written in two simple forms:

\begin{equation}\label{tr}
L(T)\approxeq\exp\biggl[\c
-\g  \Bigl(\frac{\T}{T}\Bigr)^2\biggr]
\,,
\end{equation}
or 
\begin{equation}\label{tr1}
L(T)\approxeq\ep^\c\biggl(
1-\g  \Bigl(\frac{\T}{T}\Bigr)^2\biggr)
\,.
\end{equation}

In Fig.2 we have plotted the results. As can be seen, above $2\,\T$ the discrepancy between the expression \eqref{P2} and approximations 
%____________________________________  f -2 ______________________________________________
\begin{figure}[ht]
\centering
\includegraphics[width=6.3cm]{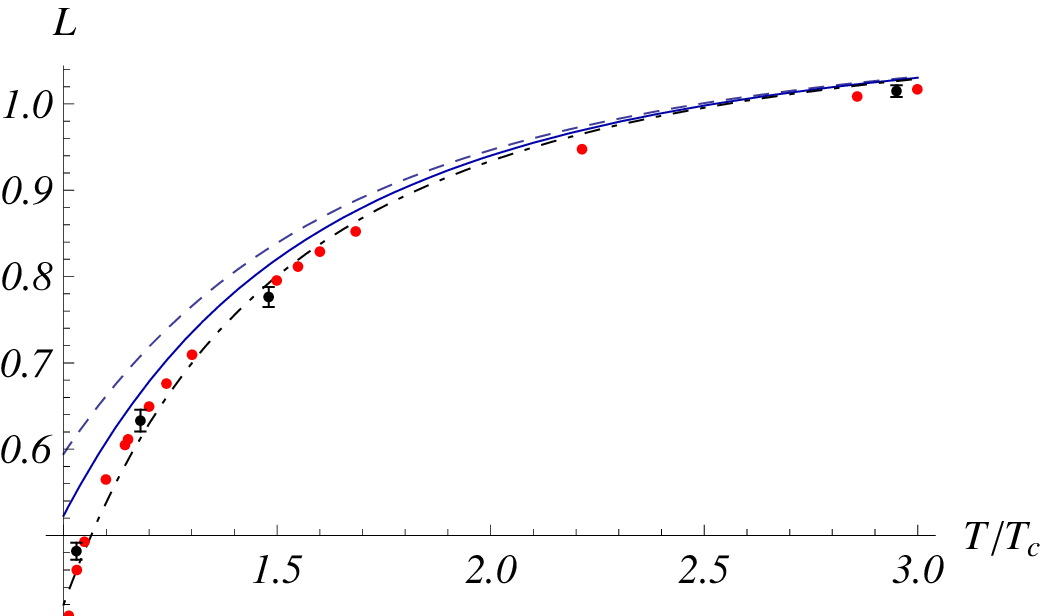}
\hfill
\includegraphics[width=6.3cm]{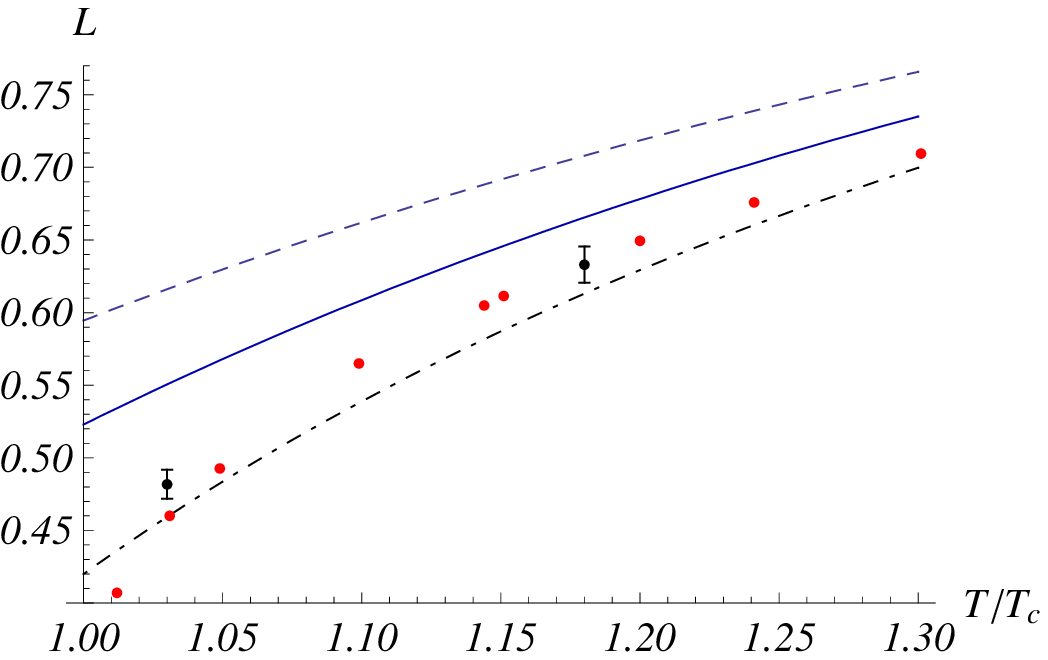}
\caption{\small{A comparison of different $L(T)$ curves for $SU(3)$ gauge theory. As in Fig.1, the solid blue curve corresponds 
to \eqref{P2} and the dots are from lattice simulations of \cite{gupta}. The blue dashed curve corresponds to the exponential law \eqref{tr}. 
The black dot-dashed curve corresponds to the power law \eqref{tr1}.  In all the cases, $\g=0.62$ and $\c=0.10$. We display error bars only 
if they are comparable to the size of the symbols. }} 
\end{figure}
%________________________________________________________________________________________
\eqref{tr}-\eqref{tr1} is negligible. At lower T the approximation \eqref{tr} (exponential law) is poor. It shows a significant deviation from the lattice. 
In particular, the discrepancy occurred at $T=1.05\,\T$ is of order 27\%. On the other hand, the agreement between the approximation 
\eqref{tr1} (power law) and the lattice is spectacular. For the temperature range $1.05\,\T \lesssim  T\lesssim 20\,\T$ the power law provides 
a reliable approximation to lattice QCD with accuracy better than 5\%! Moreover, one can use it to describe all available lattice data 
of \cite{gupta} at lower $T$. Then, the maximum discrepancy occurred at the lowest available value $T=1.012\,\T$ is of order 7\%.

It is worth noting that the exponential law has been suggested in \cite{arriola} based on a dimension-two condensate 
$\langle A^2\rangle$ \cite{2con}. Such a condensate as well as its possible links to the UV renormalon and $1/Q^2$ corrections got 
intensively discussed in the QCD literature \cite{viz}. As was first shown in \cite{1/q2}, the deformation parameter 
$\s$ of the background geometry \eqref{metric} is tied into the appearance of the quadratic corrections. It is not, therefore, surprising that we 
have recovered \eqref{tr} in our calculations.

Interestingly, the power law \eqref{tr1} is very similar to that observed for the pressure in \cite{pisa}. Indeed, for 
$T\gtrsim 1.2\,\T$ the pressure is simply $p/T^4\approx f_{\text{\tiny pert}}(1-(\T/ T)^2)$. 

%___________________________________________________________________________________________
\section{Conclusions}

In this note we have evaluated the Polyakov loop using the now standard ideas motivated by gauge/string duality. A key point is the use 
of the background metric \eqref{metric} which is singled out by the earlier works \cite{az1,az2,a-pis}. (Note that there is no need for any
free parameters except a scheme-dependent normalization constant $\c$.) The overall conclusion is that the same background metric results in a 
very satisfactory description of the Polyakov loop as well. Of course, we still have a lot more to learn before answering the question posed at 
the beginning of this note. 

\vspace{.25cm}
{\bf Acknowledgments}

\vspace{.25cm}
We would like to thank R.D. Pisarski and P. Weisz for useful discussions, and S. Hofmann for reading the manuscript. This  work  is
supported in part by DFG "Excellence Cluster'' and the Alexander von Humboldt Foundation under Grant No. PHYS0167.

%__________________                      R E F S                    ______________________

%____________________________________________________________________

\begin{thebibliography}{99}
\bibitem{s}
Alternatively, it may be fixed from the heavy quark potentials \cite{az1,white}.
\end{thebibliography}
\end{document}